\def\ANON{0} %

\if \ANON 1
\documentclass[sigconf,10pt,anonymous]{acmart}
\else
\documentclass[sigconf,10pt]{acmart}
\fi
\usepackage{algorithmic}
\usepackage{graphicx}
\usepackage{textcomp}
\usepackage{xcolor}
\usepackage{subcaption}
\usepackage{hyperref}
\usepackage{relsize}
\def\BibTeX{{\rm B\kern-.05em{\sc i\kern-.025em b}\kern-.08em
    T\kern-.1667em\lower.7ex\hbox{E}\kern-.125emX}}
    
\usepackage{xspace}

\usepackage{listings}

\usepackage{xprintlen} %

\newcommand{\rodrigo}[1]{{\color{blue}[RF: #1]}}
\newcommand{\vaastav}[1]{{\color{magenta}[VA: #1]}}
\newcommand{\ak}[1]{alokk:\textcolor{red}{#1}}

\usepackage{titlesec} %
\titlespacing*{\section}{0pt}{2mm}{1mm}  %
\titlespacing*{\section}{0pt}{2mm}{1mm}  %
\titlespacing*{\subsection}{0pt}{2mm}{1mm}  %
\titleformat{\subsubsection}[hang]
  {\normalfont\large\bfseries}
  {\thesubsubsection}{1em}{}
\titlespacing*{\subsubsection}{0pt}{2mm}{1mm}

\usepackage{setspace}  %
\usepackage{tikz}
\usetikzlibrary{external}
\usepackage{url}
\makeatletter
\g@addto@macro{\UrlBreaks}{\UrlOrds}
\makeatother

\setcopyright{none}
\renewcommand\footnotetextcopyrightpermission[1]{} %

\def\Snospace~{\S{}}

\newcommand{\todo}[1]{}
\renewcommand{\todo}[1]{{\color{red} TODO: {#1}}}
\newcommand{\fakepara}[1]{\vspace{0mm}\noindent\textbf{#1}~~}

\definecolor{figurecolor}{RGB}{22,90,220}
\definecolor{citecolor}{RGB}{198,81,19}
\hypersetup{ colorlinks=true, urlcolor=black, linkcolor=figurecolor, citecolor=citecolor, pdfstartview=FitH,
        pdftitle={Intent-based System Design and Operation},
        pdfauthor={}
        }

\begin{document}

\title{Intent-based System Design and Operation}

\if \ANON 0
  \author{Vaastav Anand\textsuperscript{\textsection}}
  \affiliation{
    \institution{Max Planck Institute for Software Systems}
    \country{Germany}}
  \email{vaastav@mpi-sws.org}

  \author{Yichen Li\textsuperscript{\textsection}}
  \affiliation{
    \institution{The Chinese University of Hong Kong}
    \country{Hong Kong}}
  \email{ycli21@cse.cuhk.edu.hk}

  \author{Alok Gautam Kumbhare}
  \affiliation{
    \institution{Microsoft}
    \country{USA}}
  \email{alokk@microsoft.com}

  \author{Celine Irvene}
  \affiliation{
    \institution{Microsoft}
    \country{USA}}
  \email{celineirvene@microsoft.com}

  \author{Chetan Bansal}
  \affiliation{
    \institution{Microsoft}
    \country{USA}}
  \email{chetanb@microsoft.com}

  \author{Gagan Somashekar}
  \affiliation{
    \institution{Microsoft}
    \country{USA}}
  \email{gsomashekar@microsoft.com}

  \author{Jonathan Mace}
  \affiliation{
    \institution{Microsoft}
    \country{USA}}
  \email{jonathanmace@microsoft.com}

  \author{Pedro Las-Casas}
  \affiliation{
    \institution{Microsoft}
    \country{Brazil}}
  \email{pedrobr@microsoft.com}

  \author{Rodrigo Fonseca}
  \affiliation{
    \institution{Microsoft}
    \country{USA}}
  \email{fonseca.rodrigo@microsoft.com}

  \fancyhead[R]{Anand et al.} %
\else
  \author{Anonymous Submission \#XXX}
\fi

\begin{abstract}
Cloud systems are the backbone of today's computing industry. Yet, these systems
remain complicated to design, build, operate, and improve. 
All these tasks require significant manual effort by both
developers and operators of these systems.
 To reduce this manual burden, in this paper we set forth a vision for achieving
holistic automation, \emph{intent-based system design and operation}. We propose
\emph{intent} as a new abstraction within the context of system design and
operation. Intent encodes the functional and operational requirements of the
system at a high-level, which can be used to automate design, implementation,
operation, and evolution of systems.
We detail our vision of intent-based
system design, highlight its four key components,
and provide a roadmap for the community to enable autonomous systems.
 \end{abstract} 
\maketitle
\begingroup\renewcommand\thefootnote{\textsection}
\footnotetext{Work done during internship at Microsoft}
\endgroup

\section{Introduction}
\label{sec:intro}

Cloud-based services are systems that are deployed in public cloud, run
continuously, and are in a constant cycle of development and operation.
 These
systems are typically distributed, have many components, and are always
evolving. They have increasingly adopted a microservice
architecture~\cite{cockcroft2016evolution,cockcroft2016microservices,uber2015soa,mazdak2017infrastructure},
where each of these components are loosely coupled, can be developed separately
using different libraries and frameworks, and scaled independently.

Full automation of cloud systems has been a long standing goal for developers
and operators.
 However, despite the persistent need, automation of the design,
building, operation, and maintenance~\cite{ganatra2023detection,ghosh2022fight}
of these systems has been elusive for multiple reasons. First, designing systems
is a long and arduous process which requires correctly handling a myriad of
complex and intertwined requirements~\cite{uber2020domain}. Second, there has
been a lack of standardization and a dearth of available tooling that can allow
developers to fully offload tasks. Third, cloud systems tend to be large and
complex which prevents easy operation and understanding as it is impossible for
any single operator to have full insight into the operational context of the
system~\cite{ganatra2023detection,ghosh2022fight}. Fourth, the environment in
which systems operate is continuously changing but the systems are not designed
to be easily reconfigurable~\cite{anand2023blueprint}.

We believe that we are now on the cusp of achieving the elusive goal of
automation. This is for two reasons. First, 
there has been a confluence of standardization and automation tools, such as 
Kubernetes for deployment, maintenance, and convergence of systems to desired
states~\cite{shanks2019k8sdesired}; 
OpenTelemetry for monitoring and
observability via logs, traces, and metrics; 
and automatic bug-finding and
verification tools that can be utilized in both development and production.
Second, the recent rise of Large Language Models (LLMs) has provided increasingly sophisticated
automatic coding and code understanding tools, and ways for operators
to interact with their system at a higher level of
abstraction.  This has the potential to reduce the significant manual effort usually required by
operators for tasks such as incident detection~\cite{srinivas2024intelligent},
incident management and
mitigation~\cite{hamadanian2023holistic,goel2024x,jiang2023xpert,ahmed2023recommending,wang2024anomaly},
and root cause
analysis~\cite{zhang2023pace,zhang2024automated,ahmed2023recommending,chen2023empowering,anand2025towards,seshagiri2024chatting}.

For cloud systems to fully embrace automation across all aspects,
we need to be able to automatically carry out actions according to the user's intent. 
To do so, we extend the idea of intent-based networking~\cite{Clemm2022}, where the network automatically configures itself to meet operators' intent, to the broader context of cloud systems by combining automation tools with the generative capabilities of LLMs.

In this paper, we introduce our vision of intent-based self-managing cloud systems.
Our goal is to enable users to specify high-level intents for the system, and have the system automatically \emph{designed, developed, and operated}. 
We envision that the creators and operators of cloud systems will be able to describe at 
a high level their functional and operational intent for the system, and automatic, intelligent
tools will be able to design, implement, and test the system, while integrating the monitoring
necessary to operate the system within desired availability, reliability, and 
safety constraints.
\section{Intent for Cloud System Design}

\begin{figure}%
\centering%
\includegraphics[width=0.48\textwidth]{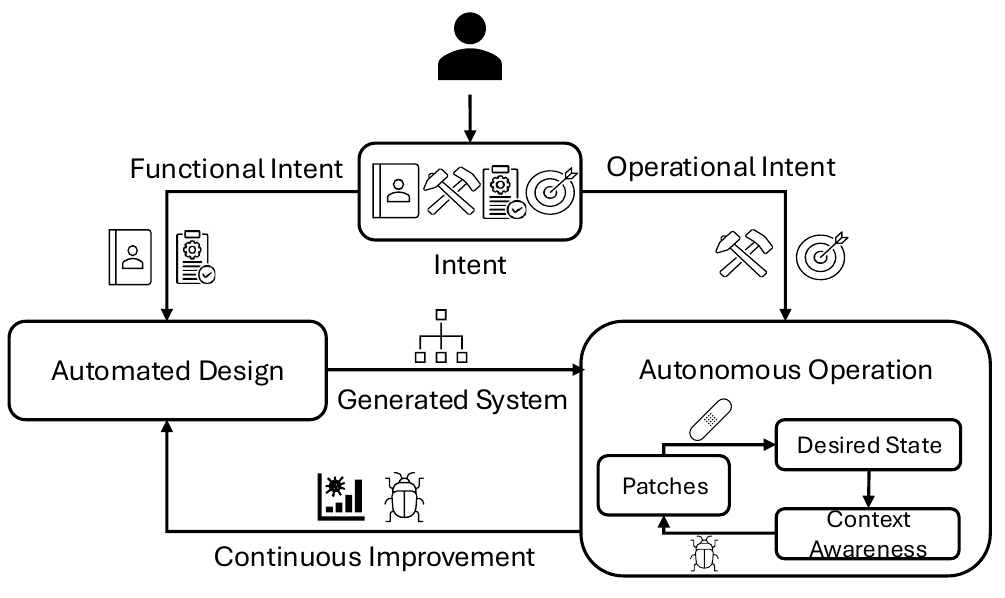}%
\vspace{-1mm}%
\caption{Intent-based cloud system components}%
\label{fig:overview}%
\if \ANON 1
\vspace{-3mm}%
\fi
\end{figure}

We introduce \emph{intent}~\cite{Clemm2022} as a high-level abstraction within the context of system design and operations.
Intent represents the potentially changing \emph{functional} and \emph{operational} requirements of the system from the user. 
\autoref{fig:overview} illustrates the proposed components for a intent-based cloud system. 
The proposed components include automated design,
automatic operation, and continuous improvement of systems.

\begin{figure}%
\begin{lstlisting}
Design a hotel reservation app as microservices. The app should allow searching for hotels near a location, search for activities near hotel, reserve and pay for a hotel room, read and write reviews about the hotels, and manage user's data.
\end{lstlisting}%
\vspace{-3mm}%
\caption{Functional Intent Example}
\label{fig:func_intent}
\end{figure}

\begin{figure}%
\begin{lstlisting}
The app should have timely responses under high load and maintain a 100ms 99th percentile latency.
\end{lstlisting}%
\vspace{-3mm}%
\caption{Operational Intent Example}
\label{fig:op_intent}
\end{figure}

We propose that there exist two broad classes of intent. 
First, \textit{functional intent} represents the feature requirements of the system. 
These include the functional requirements, security requirements, as well as design requirements.
\autoref{fig:func_intent} shows an example functional intent for a hotel reservation application.
Second, \textit{operational intent} represents the operating properties 
of the system which is used to derive the system SLA as well as metrics
and monitors for gaining detailed insights into system behavior, ways to 
detect deviations from intended behavior, and strategies for mitigating 
issues and incidents. \autoref{fig:op_intent} shows an example operational intent for an application.
To accommodate changes to systems over time, we define the desired changes in the functional and operational intent as \textit{refinement intent}.
Refinement intent represents the delta between the initial intent and the new intent and is used for automatically improving the system.

\subsection{Manifesting Intent}

Currently, developers and operators manifest intent in cloud systems as part of the
software development lifecycle (SDLC). 
SDLC is commonly divided into
six phases~\cite{abrahamsson2017agile} - (i) requirements engineering to extract desired features, (ii) prioritizing features and estimating effort, (iii) designing the system, (iv) implementing the selected design, (v) testing and productizing to ensure that the system is correct and understandable, (vi) deploying and maintaining the system.

As part of the SDLC, the functional and operational intent are extracted during the requirements engineering phase
and converted into concrete actions \emph{manually} taken by developers and operators through the other phases of the SDLC.

Instead of manifesting intent manually, we instead propose
a human-in-the-loop approach for automating different phases of the SDLC
to reduce the manual burden on developers and operators.
Our human-in-the-loop approach uses LLMs to generate concrete
actions that are then applied on the target system via automation tools.

\subsection{Challenges}

\fakepara{Reliability and Correctness.} LLMs are well known to suffer from hallucinations~\cite{huang2023survey} that lead to correctness
issues in their output.
Additionally, LLMs struggle with numerical and logical reasoning tasks leading to factually incorrect or inconsistent output. 
The output quality issue is further compounded for code generation tasks as the knowledge base of LLMs may consist of buggy, incorrect, or poorly written code.
Ensuring the reliability of their outputs requires robust verification mechanisms and careful human oversight. 

\fakepara{Explainability.} Human operators must be able to verify and understand the rationale behind the actions selected by LLMs to ensure they align with the user intent.
This requires implementing mechanisms for clear documentation, justification of actions, and validation processes to build trust and enable effective oversight. 
Failure to do so can cause unforeseen issues and monetary loss~\cite{asim2024how}. 

\fakepara{Providing accurate context.} The outputs of LLM is directly dependent on the quality of the information provided as context in the input prompt.
Cloud systems tend to be large in size spanning thousands of lines of code and documentation, generating billions of traces ~\cite{kaldor2017canopy},
and PetaBytes of logs~\cite{liu2019logzip} per day.
Extracting relevant information to serve as context for inputs to the LLM is a challenging task.

\fakepara{Action Selection.} LLMs suffer from instruction inconsistency~\cite{huang2023survey}, in which LLMs deviate from user directives.
This deviation can lead the LLM to misinterpret the user's intent and select inappropriate actions.
The problem of selecting the correct action is exacerbated by the large number of potential actions in large-scale cloud systems.

\fakepara{Dynamic Adaptation.} Certain actions selected by the LLM might require changing the system online. 
This requires the systems to dynamically adapt while still running without requiring bringing the system offline.
The system needs to have the ability to dynamically reconfigure itself and continuously update and adapt. %
\section{Intent-based self-managing cloud systems}
\label{sec:design}

In typical design and implementation of cloud systems, developers and operators manually manifest intent for four high level tasks.
In this section, we detail an LLM-based approach for automating intent manifestation for these four tasks.

\subsection{Distributed System Design}
\label{sec:system_design}

Developers often struggle with designing distributed systems because
it requires managing the complexities of multiple independent moving pieces
while ensuring the correctness, performance, and reliability of the system. To combat these issues, developers rely on principles and hints~\cite{lampson2020hints,lampson1983hints} to select an appropriate design which they then manually implement using standardized tools such as Docker and OpenTelemetry. Naturally, this is a manually arduous process.

\fakepara{Key Idea.} To alleviate the manual effort on developer, we propose using LLMs to convert the intent of the users, provided as user requirements, into concrete implementations.

\fakepara{User Requirements.} The user requirements represent the intent of the various stakeholders. 
The functional requirements and the desired architecture pattern comprise the functional intent of the system. 
The behavioral properties of the system are the operational intent of the system.

\fakepara{Requirements.} To generate a reliable, effective design of a distributed system, there are three different classes of requirements that a automatically generated system must provide. 
First, correctness guarantees, which include correctness with respect to user requirements, test suites, and formal specifications of the system. 
Second, explainability guarantees: the generated code must be understandable by humans and should provide more artifacts that can be used by developers to gain insights into the system. 
Third, performance guarantees, which may include scalability, SLOs, and absence of emergent misbehaviors~\cite{mogul2006emergent,bronson2021metastable,huang2022metastable}.

\subsubsection{Use Case: Generating Microservices}

Microservices are a pervasive design architecture commonly used for developing modern cloud systems~\cite{cockcroft2016evolution,cockcroft2016microservices}.
Due to their importance, there has been a growing interest in automating the generation and deployment of microservice systems~\cite{anand2023blueprint,ghemawat2023towards,dapr}. 
However, the effort has largely been focused on automating the generation of infrastructure components of the system rather than the business logic.

Cerulean~\cite{anand2025automated} is a human-in-the-loop system that combines the generative capabilities of LLMs to generate the business logic of the system and
then converts the business logic of the system into input specifications of Blueprint~\cite{anand2023blueprint}. 
To generate the business logic, Cerulean proposes a \emph{hierarchical generation} procedure that decomposes the system generation process
into multiple steps at different levels of abstraction of the system design process including high-level design, low-level design,
and unit-test generation. This process decomposes the functional intent and iteratively converts the intent into specific design choices of the system.

We leverage the modularity of Cerulean and extend the hierarchical generation process with two novel components to show how the \emph{hierarchical generation} process can be extended to improve the quality of the generated system. 
First, we introduce an end-to-end test-case generation component that extends the \emph{implementation generation} phase of the hierarchical generation
process to also generate end-to-end tests of the system. To do this, first the component uses LLMs to extract end-to-end use-cases from the functional intent
of the user. It then generates an implementation plan consisting of API calls to the frontend service(s) of the system using the generated interfaces in a previous phase. 
The component then uses this plan to generate a concrete end-to-end test that can be executed as a traditional go test or be compiled as a blackbox test
that can be run against the deployed system.
Second, we introduce a workload generator component that automatically generates workload processes that can be used by developers to benchmark
the deployed system. The workload generator component expects that the input operational intent includes a description of the target workload. If the description is missing, then the user is prompted to provide a description.
The component then uses this description along with the previously generated interfaces to generate a process that exercises the target workload
when executed against the deployed system. 

We believe that Cerulean's hierarchical generation process can be further extended to provide additional guarantees for the generated system
such as verification guarantees by incorporating the use and generation of formal models.

\subsection{Real-Time Context Awareness}
\label{sec:real_time_context_awareness}

\textit{Real-time context awareness} refers to the ability of the system to continuously and accurately understand its current operational state and the context in which it operates. 
\textit{This understanding is essential for the system to make informed decisions and take appropriate actions to meet the user's intent.}

\fakepara{Key idea.} To provide a cloud system the ability to continuously and autonomously comprehend its state and operational environment in alignment with the user's specified intent, we combine advanced monitoring techniques with LLMs. 
By correlating the different sources and forms of runtime data, we create a unified representation of the system's operational state. 
This representation is then connected with the system's domain knowledge (e.g., code, documentation) to generate the system \emph{context}.
As opposed to existing single-modal data approaches for analyses tasks~\cite{rosenberg2020spectrum,lee2023maat,li2022intelligent,lin2020fast},
context encapsulates and connects relevant information from various sources to provide a holistic view which can be used for further automated operations or for providing developers with relevant information for analysis tasks.

\fakepara{Context.} Context represents the current operational information of the system comprising both runtime information and domain knowledge.
Runtime information includes metrics, logs, traces, and monitors.  
Domain knowledge includes code, documentation, and operational guidelines such as troubleshooting guides. 
The real-time context awareness framework provides the intent-related contextualized and summarized knowledge for analyses tasks.

\fakepara{Requirements.} To achieve effective real-time context awareness, it is essential to have comprehensive observability that covers different aspects of the system, including performance metrics, logs and traces. 
Cloud systems generate a large amount of data. 
It is crucial to understand the intent and use it a guiding principle to filter the relevant data.

\subsubsection{Use Case: Behavior Comprehension}

\begin{figure}[ht]%
\centering%
\includegraphics[width=0.48\textwidth]{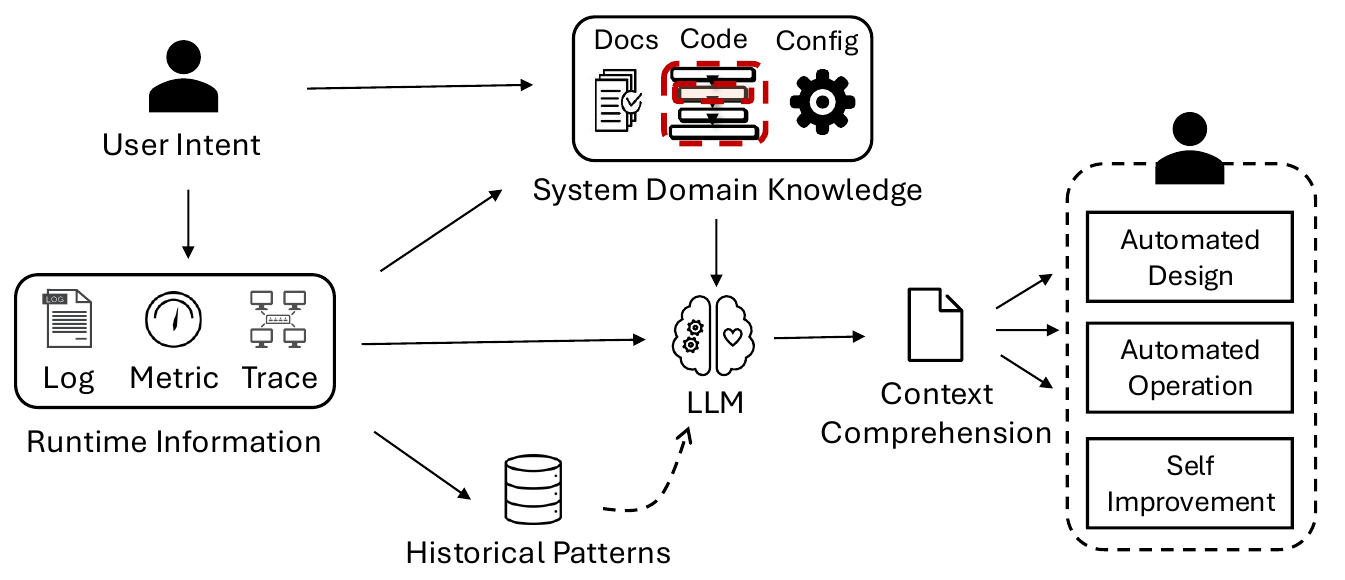}%
\caption{Real-time context awareness.}%
\label{fig:context_awareness}    
\end{figure}
We bridge runtime data with service domain knowledge based on user intent, as shown in \autoref{fig:context_awareness}. The intent is essential to determining the scope of the data that is needed. This framework allows us to extract and refine pertinent system domain knowledge, offering real-time system comprehension based on user intents.

\fakepara{Runtime Data Modeling.} We model multi-modal runtime data as loosely unified event graphs~\cite{yu2023nezha}, representing the system's current execution state for real-time retrieval and modeling based on user intent. The runtime data includes metrics, monitors, logs, and traces. 
For the metrics, we represent their events as anomalies or deviations from the normal behavior. 
For logs, we mine patterns to identify sequences that require particular attention and understanding, treating each subsequence as an event.  
Traces function as connectors between these events, delineating the interactions and dependencies among various components. 
This integrated approach allows us to create a real-time and comprehensive mapping of the system behavior.

\fakepara{System Domain Knowledge Extraction.} We extract relevant descriptions from documentation for logs and alerts that involve specific terms.
We extract the related source code for logs and integrate it into our context.

\fakepara{Comprehension Generation.} We continuously correlate and integrate context data to provide real-time system comprehension. Pattern mining and comprehension generation are triggered on-demand, avoiding unnecessary cost. The generated system comprehension serves as fundamental, shareable knowledge for users and different components.

\subsection{System Operation}
\label{sec:system_operation}

Operating a system is a complex task. Systems can be composed of many different components, 
can be built at different layers and usually evolves over time.
Despite several advances in automating different aspects of system operation~\cite{roy2024rcaagent,xie2024cloudatlas},
there is still a constant stream of failures and incidents that require human intervention.

\fakepara{Key idea.} To fully automate system operation, we translate 
operational intent provided by the user into an \emph{ops model} that is used to automate the system operation. 
We combine the model with the real-time insights from the context comprehension component, which enables the system to anticipate potential issues, automate decision-making, and execute operational tasks with minimal human intervention. 

\fakepara{Ops model.} The operational intent provides the operational properties of the system including the SLOs and SLAs.
The operation intent is translated to an operational model, \textit{ops model}, that is used to guide the system operation. 
The \textit{ops model} defines the desired state of the system and identifies and prioritizes the potential risks and vulnerabilities in the system operations to SLAs. 
It formalizes the set of observability (metrics, monitors, logs, traces) required to identify and diagnose potential issues and defines the set of mitigations and countermeasures to be taken in case of failures. 
The \textit{ops model} is not static and can evolve over time as the system evolves and the user requirements change.

\fakepara{Requirements}.To effectively operate a system based on the user's intent, it is essential to have a clear definition of the desired state, the operations, their outcomes and constraints. 
It is also important to have comprehensive understanding of the system's state and environment, including the ability to anticipate potential deviations from the desired state and to to identify and mitigate these issues.  
Furthermore, it is required an intelligent decision-making engine that can interpret the operational intent and contextual insights, enabling automated actions and proactive mitigation strategies.

\subsubsection{Use Case: Automated Incident Management}
Current cloud providers rely on human intervention guided by
troubleshooting guides (TSGs) to mitigate and resolve issues that frequently occur to their services.
Automating the execution of TSGs can significantly reduce the time to mitigation (TTM) and reduce the burden of SREs.
For example, \texttt{Llexus}~\cite{llexus} is a tool that aims to automate the execution of TSGs by using LLM agents to produce executable plans from a source TSG. 
\texttt{Llexus} uses human-generated troubleshooting guides to create executable plans. These plans are executed when new incidents occur, enabling automatic mitigation and resolution of issues in cloud services.

As noted, this current approach improves the process of automatically mitigating and resolving issues in cloud services, but it still requires human curated TSGs with high quality and coverage as input.
We envision that leveraging the \textit{ops model} along with real-time context comprehension can provide a powerful framework for automating incident management in cloud services. 
The \text{ops model}'s detailed definition of operational requirements, observability metrics, and potential mitigation, combined with the context comprehension service's 
ability to dynamically monitor and understand the system state, can enable the automatic generation of actionable instructions. 
These instructions can be used as input to \texttt{Llexus} to generate executable plans and automatically mitigate and resolve possible issues that might happen to the system.  %
\subsection{System Improvement}
\label{sec:system_improvement}

\begin{figure}[ht]%
\centering%
\includegraphics[width=0.48\textwidth]{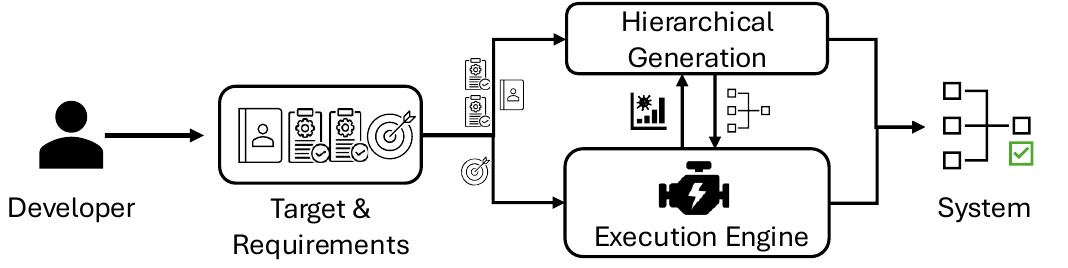}%
\caption{System improvement with Hierarchical Generation}%
\label{fig:improvement}
\end{figure}

As the system is deployed and runs, there can be deviations from intent due to several causes. 
There can be many reasons for intent violation, including unseen bugs, changes in workloads, metastable failures~\cite{bronson2021metastable}, or changes in the intent itself. 
We need the system to automatically detect such intent violations and adjust itself, either by changing configurations, fixing the code, or redesigning parts, or the whole of, the system itself.

\fakepara{Key Idea.} We detect intent violations by coupling the functional and operation intents, at different levels, with the real-time context awareness described in \autoref{sec:real_time_context_awareness} to generate the refinement intent.
We then address the refinement intent either by dynamically re-configuring the system~\cite{gagan2024} or by re-designing the system~\autoref{sec:system_design} at the desired abstraction level.
\autoref{fig:improvement} depicts the continuous improvement process that uses hierarchical generation process from Cerulean~\cite{anand2025automated} to continuously re-generate implementations of the system. 

\subsubsection{Use Case: Mitigating Metastable Failures}

Metastable failures~\cite{bronson2021metastable,huang2022metastable} are caused by a trigger that either increases the load on the system or reduces the capacity of the system to handle the load. 
These triggers are often unknown at development time; thus preventing the systems from testing and safeguarding against such scenarios.

Once these metastable failures happen in deployed systems, we use the data collected from the context-awareness and autonomous operations components to
generate trigger-scenarios which are used to reproduce the metastable failures in controlled settings of the system. 
We can then use hierarchical generation to generate new candidate designs for the system and then re-run the trigger scenarios to test if the system can avoid 
going into a metastable state while still adhering to the functional and operational intent. 
This process is repeated until a suitable design is found.
\section{Future Directions} 
\label{sec:discussion} 
The realization of fully autonomous intent-based systems necessitates research advancements in several areas. We outline the key directions to achieve this vision.

\fakepara{Manifesting Intent.} It is crucial to help users specify, refine, and understand their intent, balancing specificity, ambiguity, and user-friendliness. The goal is to minimize user burden while efficiently translating intent into service and operation models. 
It is also critical to keep the users engaged, instead of just pressing `yes' in key human-in-the-loop moments. 

\fakepara{Autonomous Design and Implementation.} Challenges include deterministic translation from models to systems, providing explanability and verifiability for generated artifacts, reasoning about design choices, optimizing for given intents and metrics, and building testable solutions. 
Mitigating LLMs hallucinations and non-determinism, integrating formal verification, and dealing with the cases in which both the generated systems and tests agree, but are wrong, are concrete areas of research.

\fakepara{Service and Operations Modeling.} These models, capturing both functional and operational requirements, form the first step in translating intent into system design. 
The challenge lies in developing comprehensive, machine-readable models that can generate and operate the system while providing explanability and verifiability.

\fakepara{Reasoning and Decision Making.} The system should incorporate intent and service context to determine the best course of action for an intent violation. 
The challenge is to develop reasoning mechanisms that can handle uncertainty, conflicting intents, and changing contexts, while providing human-consumable explanations.

\fakepara{Autonomous Operations and Self-healing.} These involve short-term mitigations and long-term fixes. A key challenge is to find the right level of intervention. 
Continuous learning and adaptation, an integral part of the system, presents challenges including representation, guard-rails, reliability of operations, and security.

\fakepara{Software Engineering Processes} must evolve to support intent-based systems. Traditional methodologies like Agile will need to incorporate autonomy to support the design, implementation, testing, and deployment of such systems. %
\section{Conclusion} 
\label{sec:conclusion}
We presented a vision for an intent-based cloud system design and operation that aims to enable fully autonomous systems that can understand, design, operate, and improve themselves based on user intent. 
We believe that such a grand vision would require multiple research communities to explore solutions in tandem to address these challenges and lead to a new class of systems and services equipped with improved productivity, reliability, and maintenance. 
\bibliographystyle{abbrv}
\bibliography{paper}

\label{page:last}
\end{document}